\documentclass{ws-p8-50x6-00}
\def\lsi{\raise0.3ex\hbox{$<$\kern-0.75em\raise-1.1ex\hbox{$\sim$}}}
\def\gsi{\raise0.3ex\hbox{$>$\kern-0.75em\raise-1.1ex\hbox{$\sim$}}}
\newcommand{\lsim}{\mathop{\lsi}}
\newcommand{\gsim}{\mathop{\gsi}}

\begin{document}

\title{Non-equilibrium dynamics\\ of hot Abelian Higgs model}

\author{A. Rajantie}

\address{Centre for Theoretical Physics,\\
University of Sussex,\\
Brighton BN1 9QH, 
\\United Kingdom\\E-mail: a.k.rajantie@sussex.ac.uk}


\maketitle

\abstracts{
The real-time dynamics of 
finite-temperature gauge theories can be approximated,
to leading-order accuracy in the coupling constants,
by a classical field theory with the hard thermal loop
Lagrangian.
I show how this approach can be used in numerical lattice
simulations to study dynamics of the Abelian Higgs model in or
slightly out of equilibrium.
}

\section{Hard thermal loops}
Depending on its details, the electroweak phase transition may
explain the observed baryon asymmetry in the 
Universe.\cite{ref:baryogenesis}
Although static equilibrium properties such as the phase structure 
of the theory have been determined
to high accuracy with numerical lattice simulations,\cite{ref:su2higgs}
much less is known about
non-equilibrium dynamics, or even 
real-time correlators in equilibrium.
The reason is that 
to calculate any real-time correlator,
one needs to evaluate
a Minkowskian path integral,
which cannot be done using Monte Carlo methods, and perturbation
theory breaks down because of infrared problems.

Assuming that 
the fields are initially in thermal equilibrium, the
occupation number of the soft modes (with momentum $k\lsim gT$) is large, 
and they can be approximated by classical fields.\cite{ref:classical}
The hard modes ($k\gsim T$) have a small
occupation number and must be treated as quantum fields, 
but perturbation theory works well for them, and they can
be integrated out
perturbatively. 
This is done by choosing a lattice cutoff $\delta x=1/\Lambda$ with
$gT\ll\Lambda\ll T$, calculating
one-loop correlators both in the full theory and in the lattice
theory and determining the effective Lagrangian by matching
the results.
Because $\Lambda$ acts as an infrared cutoff, this construction is free 
from infrared problems, 
and the expansion parameter is $g^2$, not $g^2T/m$ as usually
is the case in perturbative calculations.
In practice, since
the full lattice result\cite{ref:nauta} is very cumbersome and therefore
unsuitable for numerical simulations, it is easier to match only
static correlators, which gives
a small error to the results.

In this way, an effective theory can be constructed for the soft modes,
and the dynamics of this effective theory is well described by
classical equations of motion.
This approach has previously 
been used to measure the sphaleron rate of hot SU(2) gauge 
theory.\cite{ref:pointparticles,ref:vlasov}
I will argue that it 
can also be used for simulating non-equilibrium dynamics of
phase transitions.

In the case of the Abelian Higgs model,
the effective Lagrangian is\cite{ref:oma}
\begin{eqnarray}
{\cal L}_{\rm HTL}&=&-\frac{1}{4}F_{\mu\nu}F^{\mu\nu}
-\frac{1}{4}m_D^2
\int\frac{d\Omega}{4\pi}F^{\mu\alpha}
\frac{v_\alpha v^\beta}{(v\cdot\partial)^2} F_{\mu\beta}\nonumber\\
&&+(D_\mu\phi)^*D^\mu\phi
-m_T^2\phi^*\phi-\lambda(\phi^*\phi)^2,
\label{equ:lagrHTL}
\end{eqnarray}
where $m_D^2=\frac{1}{3}e^2T^2-\delta m_D^2$ and
$m_T^2=m^2+(e^2/4+\lambda/3)T^2-\delta m_T^2$.
The mass counterterms are\cite{ref:lainejama}
$\delta m_D^2=e^2\Sigma T/4\pi\delta x$ and 
$\delta m_T^2=(3e^2+4\lambda)\Sigma T/4\pi\delta x$.
The integration 
is taken over the unit
sphere of velocities $v=(1,\vec{v})$, $\vec{v}^2=1$.
The extra terms in the Lagrangian (\ref{equ:lagrHTL})
depend only on the couplings and the temperature, as long
as the high-temperature approximation $T\gg m$ is valid. 

The equations of motion can be derived from Eq.~(\ref{equ:lagrHTL}), and are
\begin{eqnarray}
\partial_\mu F^{\mu\nu}&\!=\!&m_D^2\int\!\frac{d\Omega}{4\pi}
\frac{v^\nu v^i}{v\cdot\partial}E^i-2e{\rm Im}
\phi^*D^\nu\phi,
\nonumber\\
D_\mu D^\mu\phi&\!=\!&-m_T^2\phi-2\lambda(\phi^*\phi)\phi.
\label{equ:nonlf}
\end{eqnarray}
Because of the derivative in the denominator, the gauge field equation of
motion is non-local.

\section{Local formulation}
\label{sec:local}
To make numerical simulations feasible, one needs a local formulation
for the theory. 
The most straightforward approaches involve describing the hard modes
by a large number of charged point particles,\cite{ref:pointparticles}
or by the phase-space distribution of these 
particles.\cite{ref:vlasov,ref:iancu}
In practice, both formulations lead to a 5+1-dimensional 
field theory.
However, in the Abelian case, one can integrate
out one of the dimensions, thus obtaining a 4+1-dimensional theory that
is completely equivalent with the others.\cite{ref:oma}  It consists of two
extra fields $\vec{f}(t,\vec{x},z)$ and $\theta(t,\vec{x},z)$,
where $z\in [0,1]$. In the temporal gauge, they satisfy the equations of
motion
\begin{eqnarray}
\partial_0^2{\vec{f}}(z)&=&z^2\vec{\nabla}^2\vec{f}+m_Dz\sqrt\frac{1-z^2}{2}
\vec\nabla\times\vec{A},\nonumber\\
\partial_0^2{\theta}(z)&=&z^2\vec{\nabla}\cdot\left(
\vec{\nabla}\theta-m_D\vec{A}\right),\nonumber\\
\partial_0^2{\vec{A}}&=&-\vec\nabla\times\vec\nabla\times\vec{A}
-2e{\rm Im}\phi^*\vec{D}\phi\nonumber\\&&
+m_D\int_0^1dzz^2\left(
\vec\nabla\theta
-m_D\vec{A}+
\sqrt{\frac{1-z^2}{2z^2}}\vec\nabla\times\vec{f}
\right).
\label{equ:eoms}
\end{eqnarray}

With these equations, it is possible to calculate
any real-time correlator at finite 
temperature. One simply takes a large number of initial configurations
from the thermal ensemble with the probability distribution $\exp(-\beta H)$,
using the Hamiltonian\cite{ref:oma} corresponding to the equations of
motion (\ref{equ:eoms}),
and evolves each configuration in time, measuring 
the correlator of interest. The average over the initial configurations
gives the ensemble average of the correlator.

\section{Simulations}
In addition to the standard lattice discretization, 
the dependence on the new coordinate $z$ needs to be
discretized as well.\cite{ref:oma} 
A convenient way is to define the canonical momenta 
$\vec{F}=\partial_0\vec{f}$ and
$\Pi=\partial_0\theta$ and to expand both the fields and the momenta 
in terms of Legendre 
polynomials:
\begin{eqnarray}
\vec{f}^{(n)}&=&\int_0^1dzz\sqrt\frac{2}{1-z^2}P_{2n}(z)\vec{f}(z),
\quad
\theta^{(n)}=\int_0^1dzP_{2n}(z)\theta(z),
\nonumber\\
\vec{F}^{(n)}&=&\int_0^1\frac{dz}{z}\sqrt\frac{2}{1-z^2}P_{2n}(z)\vec{F}(z),
\quad
\Pi^{(n)}=\int_0^1dzP_{2n}(z)\Pi(z).
\label{equ:defLeg}
\end{eqnarray}

The Hamiltonian can be written in terms of these Legendre modes, 
and the ensemble of initial configurations
can be generated using standard Monte Carlo techniques. 
This can be done in two steps:
\begin{itemize}
\item[(i)] The Hamiltonian is Gaussian in the hard fields
$\vec{f}$, $\vec{F}$, $\theta$ and $\Pi$, and they can therefore be
integrated out analytically. This leads to the Hamiltonian of the
ordinary classical Abelian Higgs model, 
with an extra Debye screening
term for the electric field. This classical Hamiltonian 
can be used to
generate the initial configuration for the soft modes.
\item[(ii)] Given the soft configuration from step (i),
the hard field configuration can be generated very effectively,
since the hard Hamiltonian is Gaussian.
\end{itemize}
After the initial configurations have been generated, any real-time
correlator can be measured as was explained in the end of 
Sec.~\ref{sec:local}.

\section{Non-equilibrium dynamics}
In addition to equilibrium real-time correlators,
the formulation presented here can also be used to study
non-equilibrium dynamics, provided that
the hard modes remain close enough to the equilibrium. This is
plausible in a phase transition, since the phase of a system is 
a property of the long-wavelength modes only. 
Near the phase transition, in both phases, all the masses
are suppressed by powers of the coupling constant $g$ relative to 
the temperature, and the high-temperature
approximation can be used. In this
approximation, the phase of the system does not enter
the results of the one-loop diagrams,
i.e.~the distribution of the hard modes is indeed
the same in both phases, within the accuracy of our approach.
Thus, there is
no reason for the hard modes to fall out of equilibrium
during the transition.

A natural way of studying a phase transition would be to start from
thermal equilibrium in the Coulomb phase, and decrease the temperature 
so that the system undergoes a transition to the Higgs phase.
This requires a mechanism for changing the temperature, and
in practice, it is easier to keep the temperature constant
and change the parameters,
such as the mass of the Higgs field, instead. In fact, when $\lambda\ll e^2$,
even that is not necessary.
The transition is of first order, and if one thermalizes
the system initially to the metastable Coulomb phase below $T_c$,
bubbles of the Higgs phase nucleate during the time evolution, and the phase
transition takes place. Assuming that the latent heat is small enough,
the temperature does not change significantly. In this way,
many non-equilibrium properties of the phase transition can be studied 
non-perturbatively.

\section*{Acknowledgments}
I would like to thank M.~Hindmarsh for collaboration on this topic.

\end{document}